\documentclass{aa}
\usepackage{graphicx}
\begin{document}

\authorrunning {Elisabetta Memola, Christian Fendt and Wolfgang Brinkmann}
\titlerunning {Thermal X-ray spectra}
 	
\title{Theoretical thermal X-ray spectra of relativistic MHD jets}

\author{Elisabetta Memola
          \inst{1,\hspace{-0.1cm}}
          \thanks{Current address: 
    Italian Space Agency - Science Data Center, c/o ESA-ESRIN,
via Galileo Galilei, 00044 Frascati, Italy},
	  Christian Fendt
 	  \inst{1,2},
          \and
          Wolfgang Brinkmann\inst{3}
          }
    \offprints{Elisabetta Memola \\ \email{memola@asdc.asi.it}}     
   \institute{Astrophysikalisches Institut Potsdam,
	      An der Sternwarte 16, 14482 Potsdam,
	      Germany
              \\
 	      \email{ememola@aip.de, cfendt@aip.de}
         \and
              Universit\"at Potsdam, Institut f\"ur Physik, 
              Am Neuen Palais 10, 14469 Potsdam, Germany
         \and
              Centre for Interdisciplinary Plasma Science,
              Max-Planck-Institut f\"ur extraterrestrische Physik,\\
	      Giessen\-bach\-strasse, 85740 Garching, Germany \\
              \email{wpb@mpe.mpg.de}
             }

   \date{Received 18 December 2001 / Accepted 4 February 2002}
   
\abstract{
Highly relativistic jets are most probably driven by strong magnetic fields
and launched from the accretion disk surrounding a central black hole.
Applying the jet flow parameters (velocity, density, temperature)
calculated from the magnetohydrodynamic (MHD) equations, 
we derive the thermal X-ray luminosity along the inner jet flow
in the energy range $0.2-10.1$\,keV. 
Here, we concentrate on the case of Galactic microquasars 
emitting highly relativistic jets.
For a $5\,{\rm M}_{\odot}$ central object and a jet mass flow rate of 
$\dot{M}_{\rm j} = 10^{-8}\,{\rm M}_{\odot}{\rm yr}^{-1}$
we obtain a jet X-ray luminosity 
$ L_{\rm X} \approx 10^{33}\,{\rm erg\,s}^{-1}$.
Emission lines of Fe XXV and Fe XXVI are clearly visible.
Relativistic effects such as Doppler shift and boosting were considered for
different inclinations of the jet axis.
Due to the chosen geometry of the MHD jet the inner X-ray emitting part 
is not yet collimated.
Therefore, depending on the viewing angle, the Doppler boosting does not
play a major role in the total spectra.
%
%
   }

   \maketitle


\section{Introduction}
Microquasars (Mirabel \& Rodr$\acute{\rm {\i}}$guez \cite{mirabel})
are Galactic X-ray binaries where the three basic ingredients of 
quasars are found --
a central black hole, an accretion disk and relativistic jets.
Jets are thought to be driven by magnetohydrodynamic (MHD) mechanisms
(Blandford \& Payne \cite{bp82};
Camenzind \cite{ca86})
triggered by the interaction of those three components, 
although the jet formation process is not yet fully understood 
(e.g.~Fendt \cite{fendt}).
Some microquasars are superluminal sources, 
e.g.~GRS\,1915+105 at a distance of $7-12$\,kpc 
(Fender et al.~\cite{fender})
with a central mass of about
$14\,{\rm M}_{\odot}$ (Greiner et al.~\cite{greiner}).

Fendt \& Greiner (\cite{fg01}, FG01) presented solutions of the MHD 
{\em wind 
equation} in Kerr metric with particular application to microquasars.
These solutions provide the flow dynamics along a prescribed poloidal
magnetic field line. 
FG01 found temperatures up to more than $10^{10}$\,K in the innermost part 
of the jet 
proposing that thermal X-rays might be emitted from this region.
Here, we calculate the thermal spectrum of such an optically thin 
jet flow taking into account one of the solutions of FG01
and considering relativistic Doppler shifting and boosting as well as
different inclinations of the jet axis to
the line-of-sight (l.o.s.).
A similar approach was undertaken by Brinkmann \& Kawai 
(\cite{bk2000}, BK00) who have been modeling the two 
dimensional hydrodynamic outflow of SS 433
applying various initial conditions.
However, they do not consider relativistic effects such as Doppler 
boosting in their spectra.

   \begin{figure*}
   \centering	 
   \includegraphics[width=4.4cm]
    {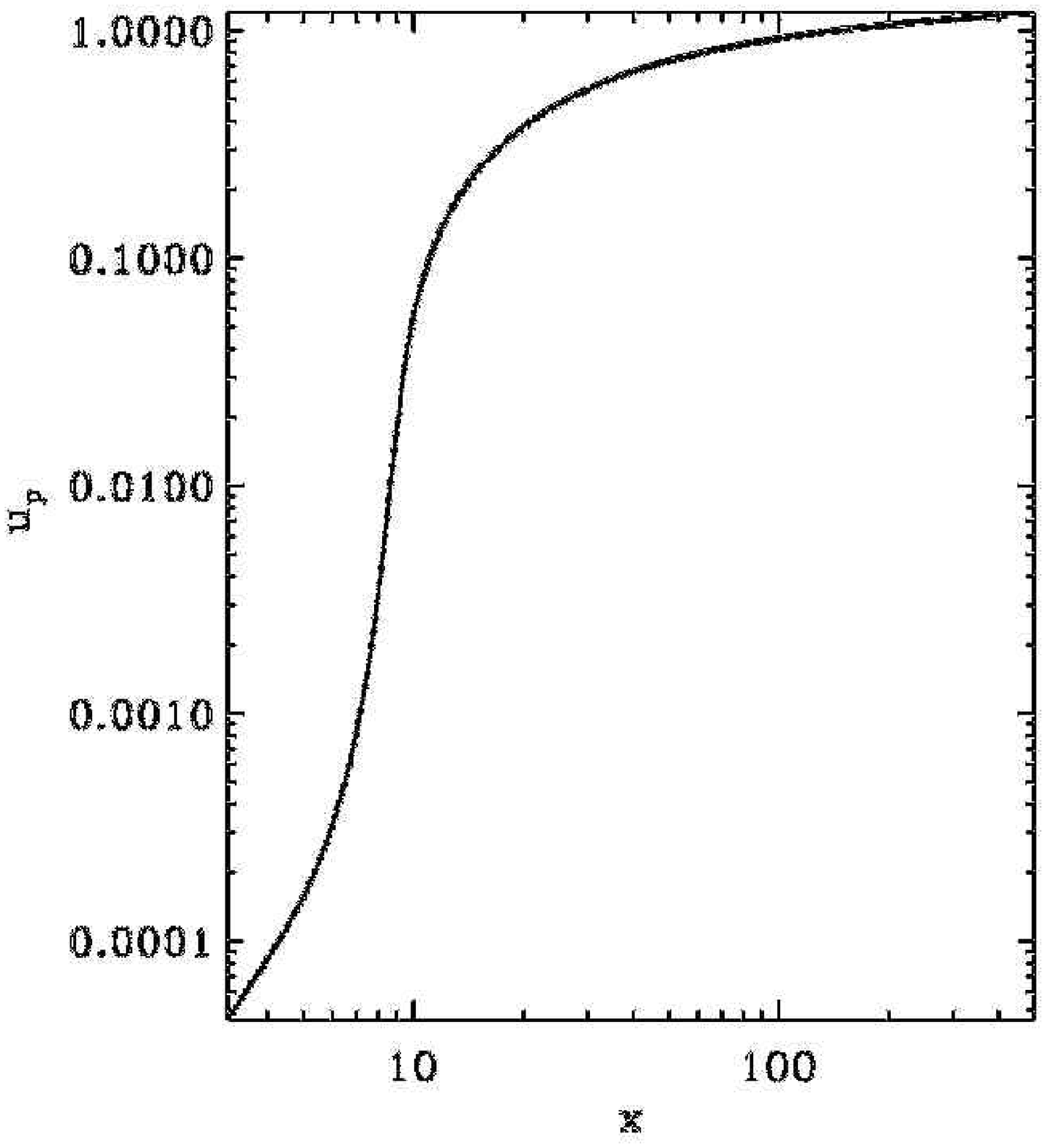}
   \includegraphics[width=4.3cm]
    {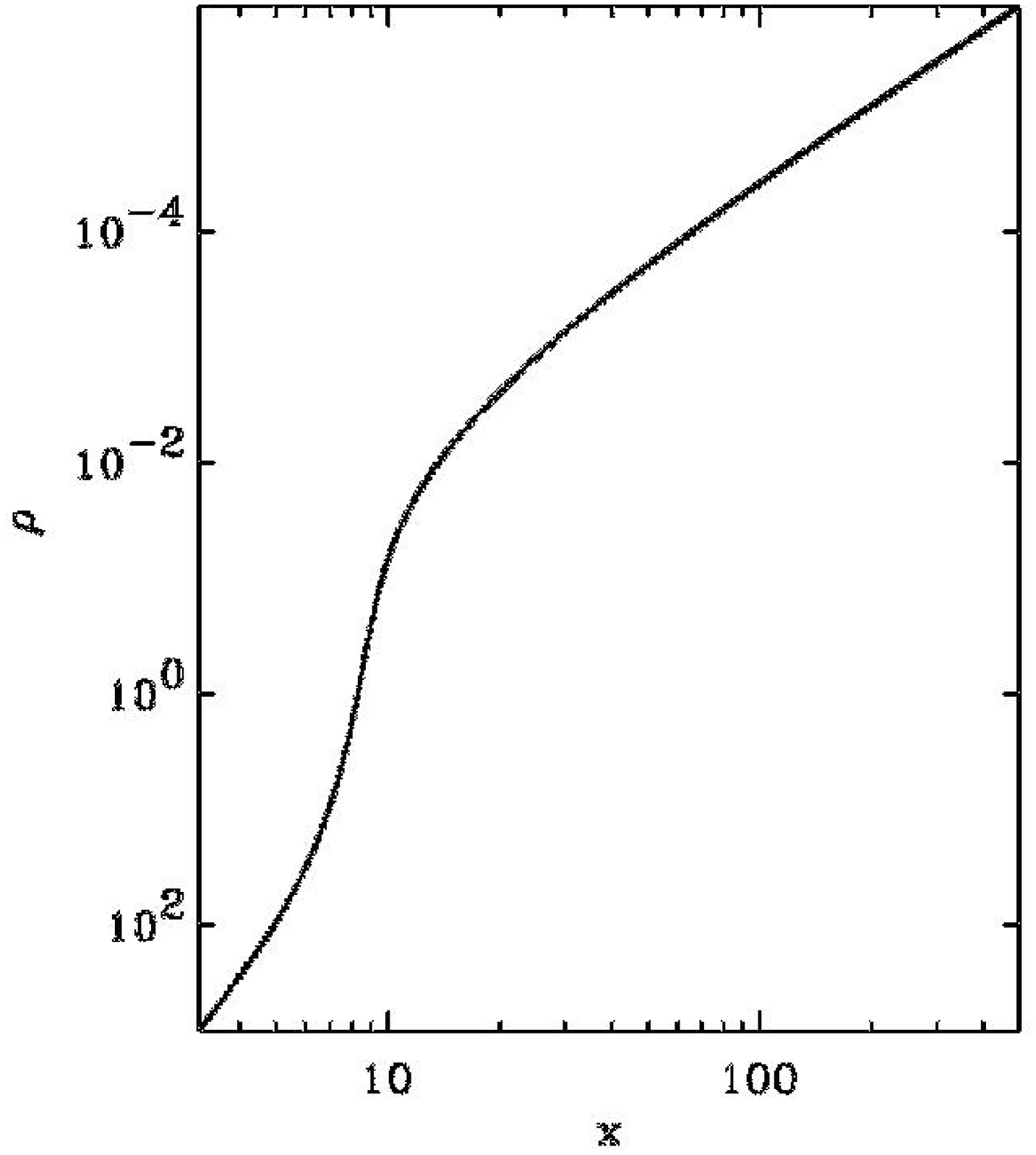}	
   \includegraphics[width=4.4cm]
    {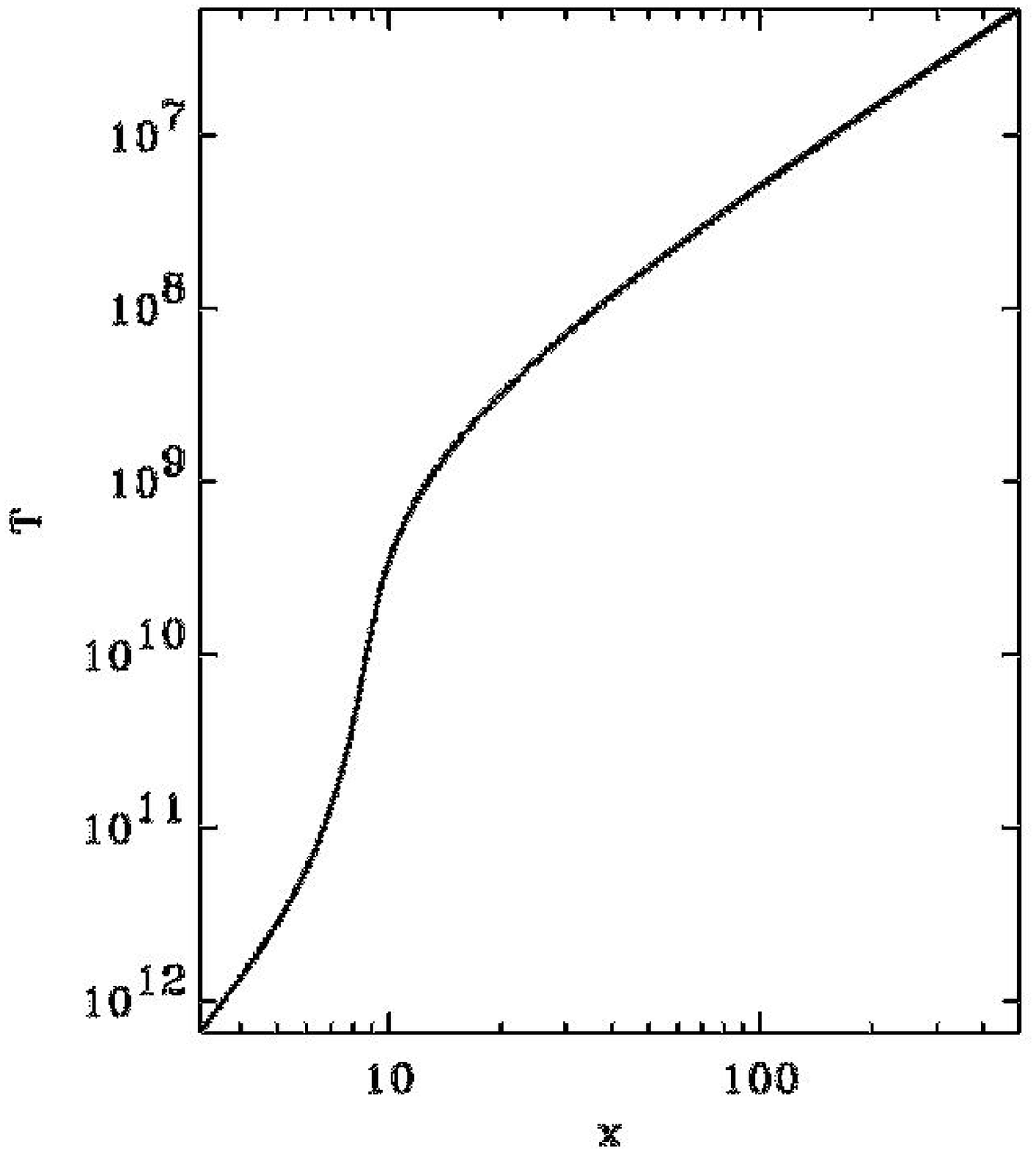}	 
   \includegraphics[width=4.4cm]
    {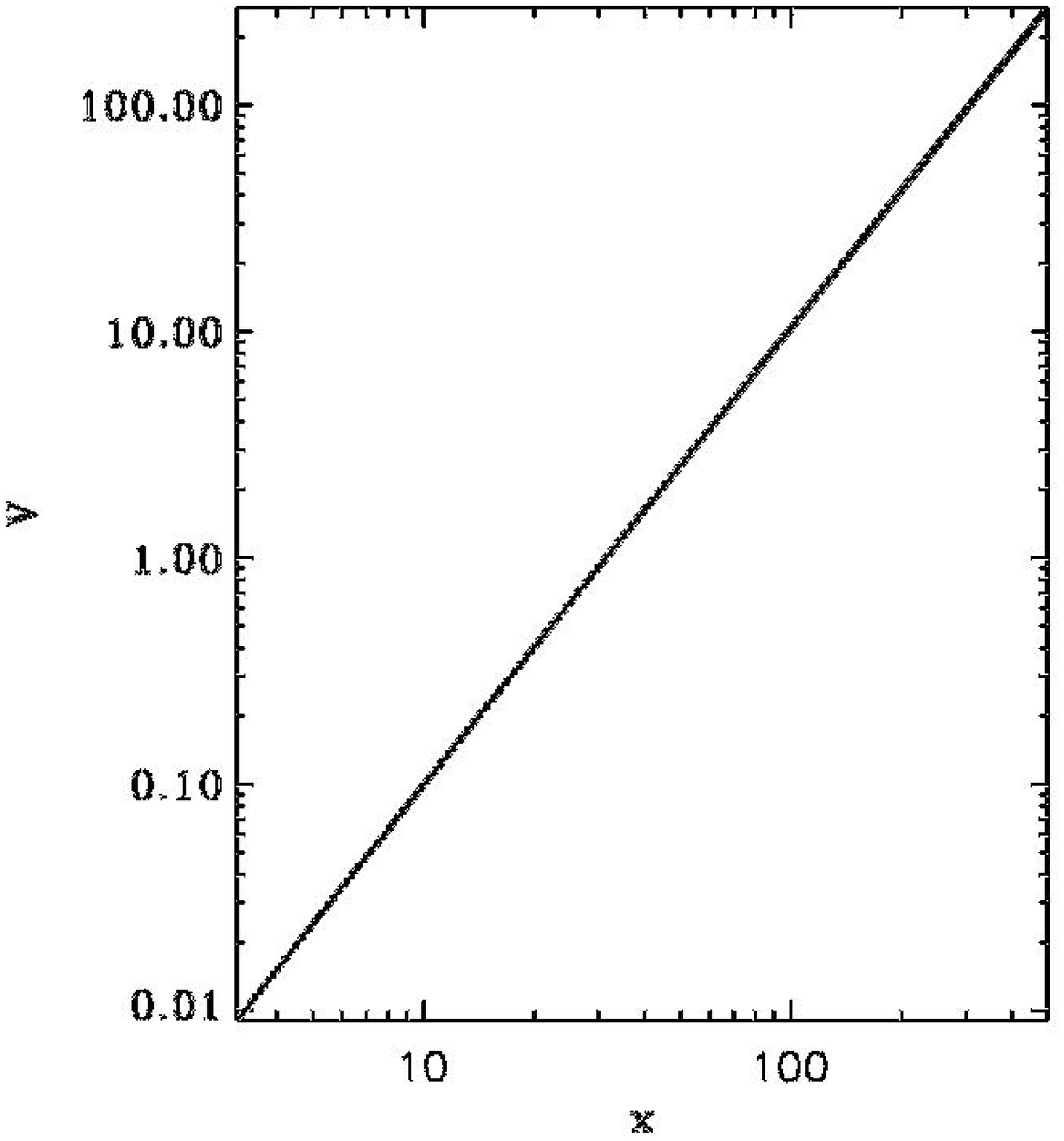}
   \caption{{Dynamical parameters of the MHD jet (see FG01).
   Shown is the radial dependence of the properly normalized 
   poloidal velocities $u_{\rm p}(x) = \gamma v_{\rm p}/c$, 
   particle densities $\rho (x)$, 
   temperatures $T(x)$ (in K), 
   and size of the emitting volumes $V(x)$ 
   (from {\it left} to {\it right}) 
   along the chosen magnetic field line.
   For the calculations in this paper we apply a central mass
   of $5\,{\rm M}_{\odot}$ 
   and a jet mass flow rate of 
   $ \dot{M}_{\rm jet} = 10^{-10}\,{\rm M}_{\odot}{\rm yr}^{-1}$.
   The units are therefore 
   $r_{\rm g} = 7.4 \times 10^5$\,cm for all length scales, 
   $r_{\rm g}^3 = 4.1 \times 10^{17}$\,cm$^3$ for the volumes,
   and $4.31 \times 10^{16}$\,cm$^{-3}$ for the particle
   densities.
   Note that the jet injection point is located at 
   $R_{\rm i} = 8.3\,r_{\rm g} $
   with a gas temperature of $T_{\rm i} = 10^{10.2}$\,K.
   In this solution for the MHD wind equation, the poloidal velocity 
   saturates to a value of $u_{\rm p} = 2.5$ beyond $x\simeq 10^8$.
   The flow is weakly collimated reaching a half opening angle of
   $70\degr$ at about $x=250$.
   }}
   \end{figure*}


\section{The model}
The axisymmetric, stationary and ideal MHD wind solution provides
the density, velocity and temperature for each volume element along the 
field.
Prescribing the jet mass flow rate $\dot{M}_{\rm j}$ together with
the shape of the field line, 
these solutions give a unique set of parameters of the flow
defined by the regularity condition across the magnetosonic points
(see FG01 for details).

The calculated dynamical parameters are our starting point
to obtain the X-ray spectra of the jet.
Here, we refer to the solution S3 of FG01 obtained for a collimating field 
line $z(x) = 0.1 (x-x_0)^{6/5} $, 
$x$ being the normalized cylindrical radius, 
$x_0$ the foot point of the field line at the equatorial plane,
and $z$ the height above the disk.
Length scales are normalized to the gravitational radius 
$r_{\rm g} =7.4 \times 10^5 $\,cm $(M/5\,{\rm M}_{\odot})$. 
For completeness, 
we show the radial profiles of poloidal velocity,
density, temperature, and emitting volume along the field line
in Fig.~1.

The jet geometry consists of nested collimating conical magnetic surfaces
with sheets of matter accelerating along each surface.
The sheet cross section becomes larger for larger distances from the 
origin.
The distribution of the 5000 volume elements along the jet is such that
velocity and density gradients are small within the volume.
We have 63 volumes in $\phi$ direction defining an axisymmetric torus
(i.e. 5000 tori along the magnetic surface).

We distinguish two parts of the inner jet flow.
One is for a temperature range $T = 10^{6.6}-10^9$\,K, where we
calculate the optically thin continuum (bremsstrahlung) 
and the emission lines.
The other is for $T = 10^9-10^{12}$\,K, where only brems\-strahlung 
is important.
Any pair processes are neglected and no ($e^-e^-$)-bremsstrahlung
will be considered, although that might be dominant at the highest 
temperatures.
Such unphysically high temperatures are to a certain degree caused by the
use of a non-relativistic equation of state. Employing a relativistically
correct equation of state (Synge \cite{synge}) 
one would expect gas temperatures 
an order of magnitude lower (Brinkmann \cite{brink}).
These temperatures belong to the intermediate region between disk 
and jet.
The injection radius,which is, in fact, 
the boundary condition for the jet flow, 
is located at $R_{\rm i} = 8.3\,r_{\rm g} $
and at a height above the disk (and the foot point of the field line)
of $0.74\,r_{\rm g}$. 
For the chosen MHD solution the temperature at this point is
$T_{\rm i} = 10^{10.2}$\,K.
With $R_{\rm i} \simeq 6 \times 10^6$\,cm $(M/5\,{\rm M}_{\odot})$
we investigate a region of about 
$2.5 \times 10^{-5}$ $(M/5\,{\rm M}_{\odot})$\,AU.
 
Having determined the emissivities of single volume elements, 
these can be put together obtaining a rest-frame 
spectrum where any motion is neglected.
However, the knowledge of the MHD wind velocities allows us to 
determine 
the Doppler shift of the spectral energies and 
the boosting of the luminosity for each volume.
We finally obtain a total spectrum of the inner jet considering
these effects in a {\em differential} way for each volume element.
The final spectra of course depend also from the jet inclination.

We emphasize that our approach is not (yet) a {\em fit} to 
certain observed spectra.
In contrary, for the first time, for a jet flow with characteristics 
defined
by the solution of the MHD wind equation, we derive its X-ray spectrum.
Our free parameters are 
the mass of the central object $M$ defining the length scales, 
the jet mass flow rate $ \dot{M}_{\rm j}$ 
and the shape of the poloidal field lines.
In the end, from the comparison of the theoretical spectra with
observations, we expect to get information about the internal 
magnetic structure of the jet close to the black hole and the jet
mass flow rate.

   \begin{figure*}
   \centering	 
   \includegraphics[width=4.4cm]
    {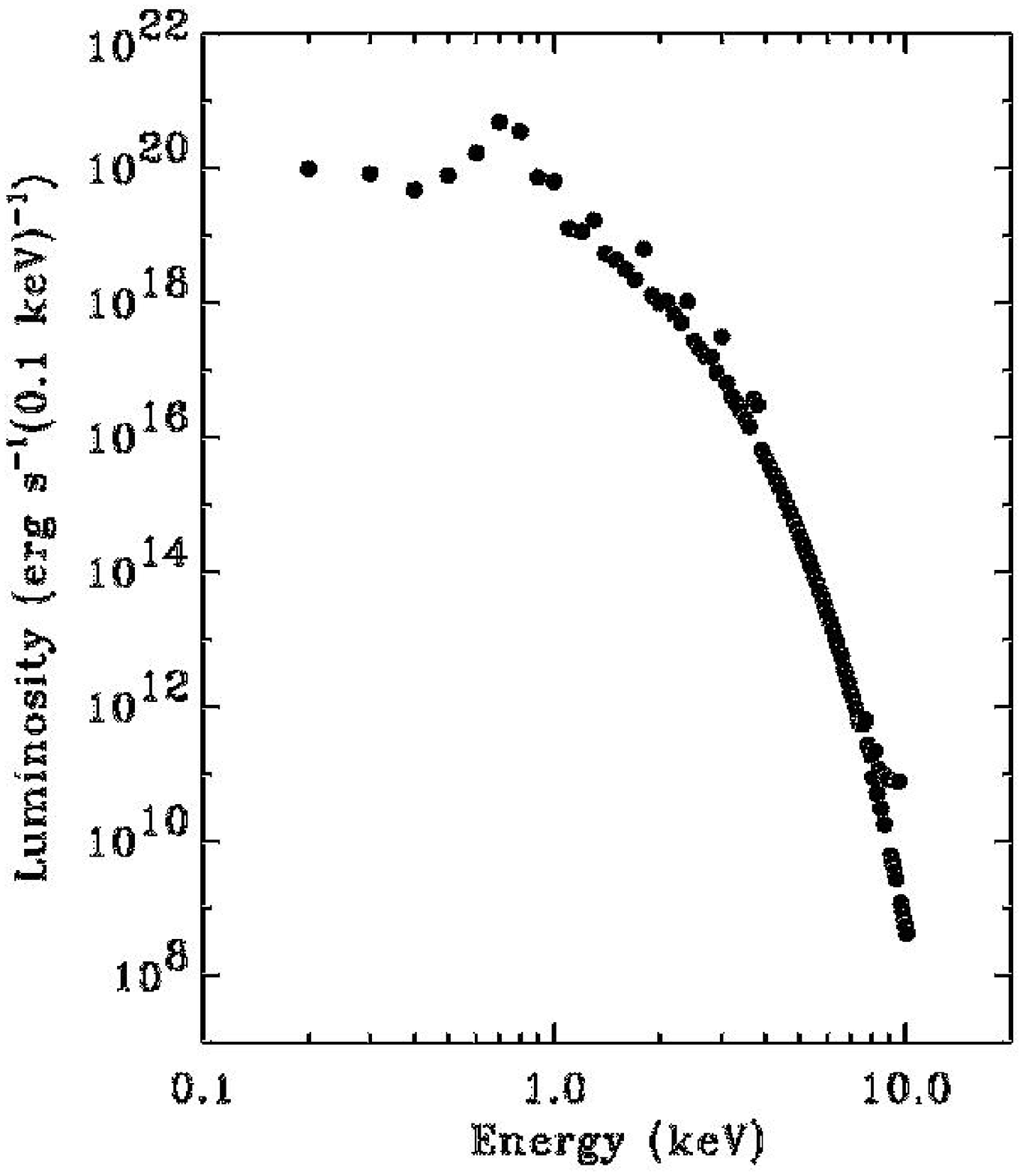}
   \includegraphics[width=4.3cm]
    {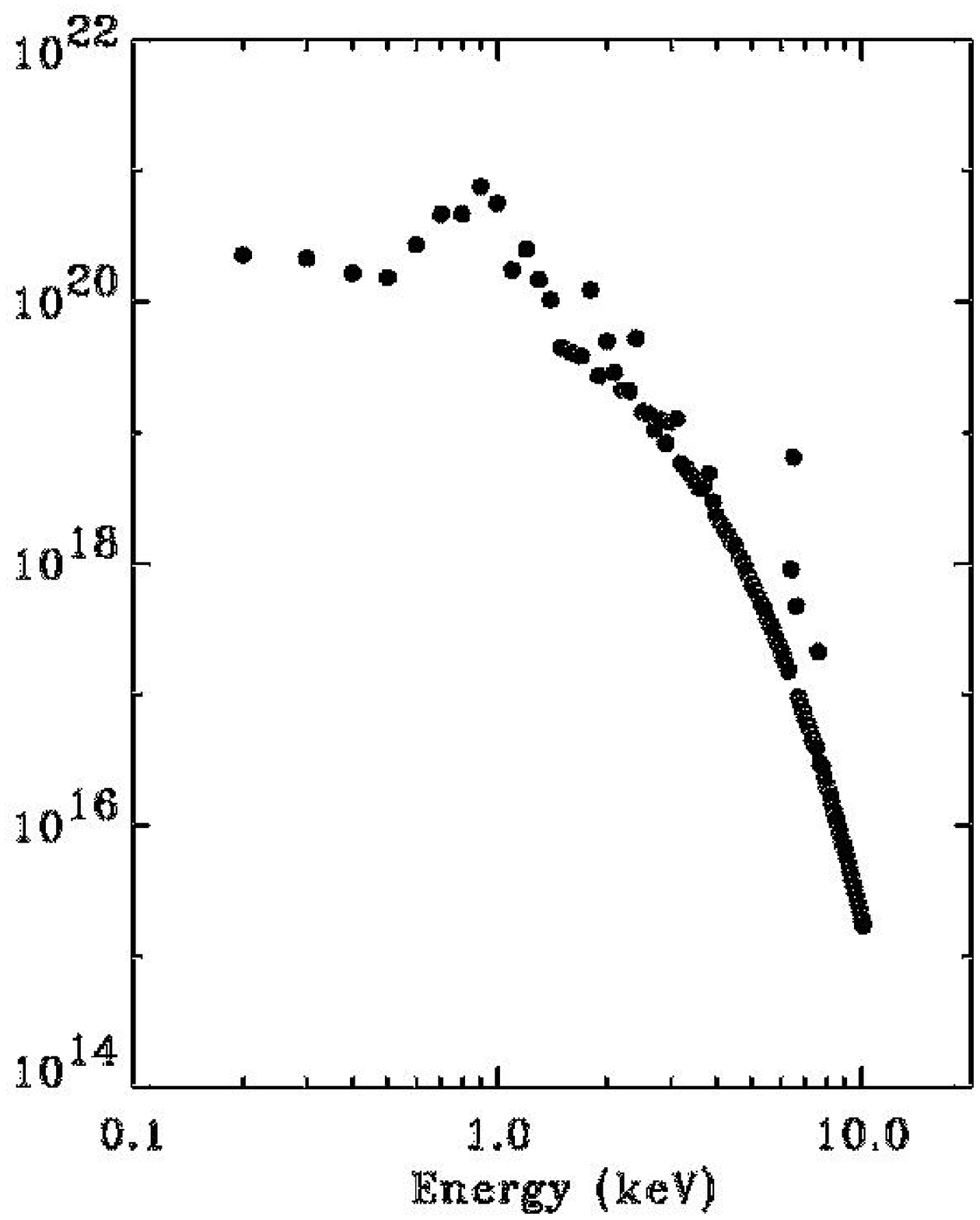}
   \includegraphics[width=4.2cm]
    {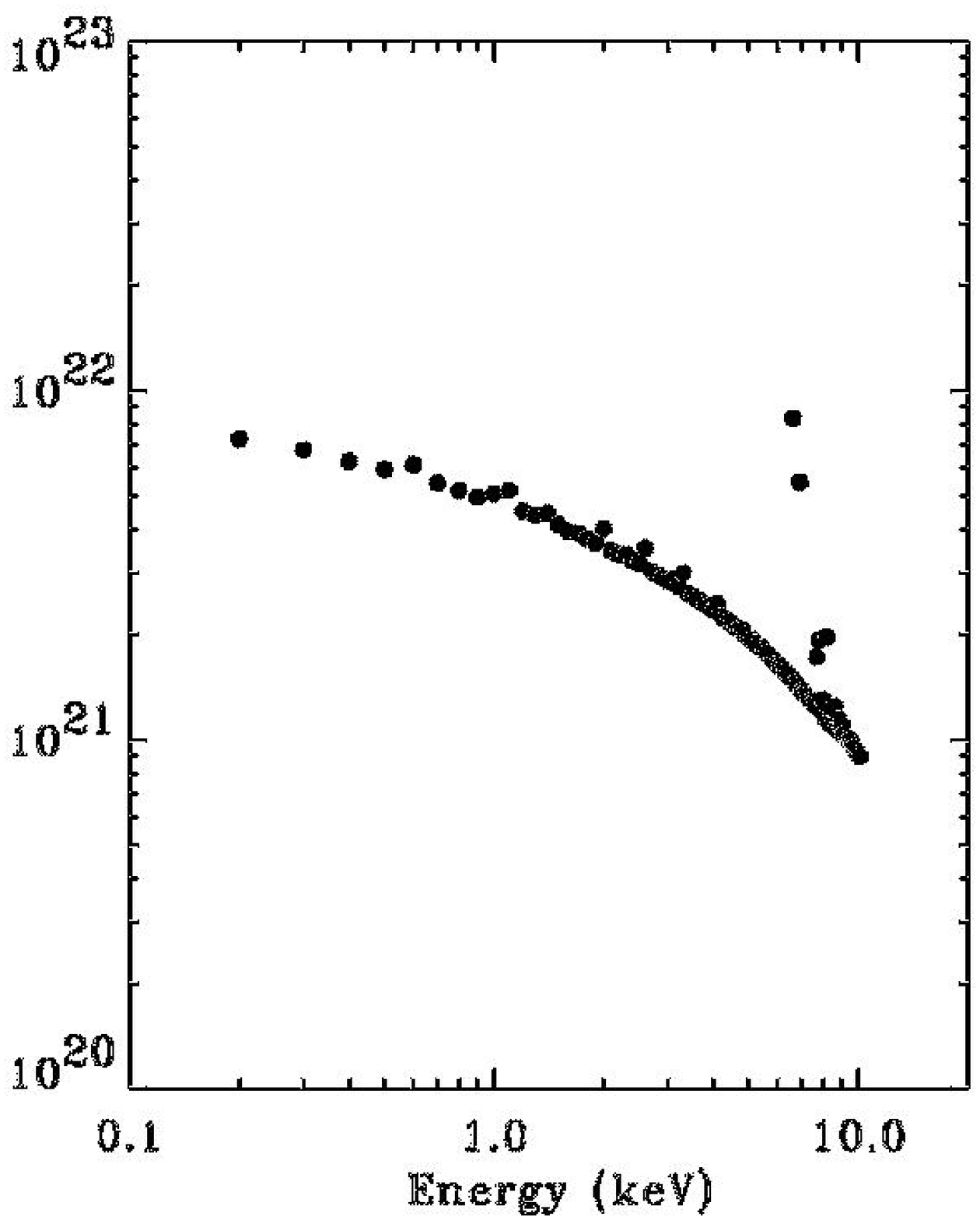}
   \includegraphics[width=4.1cm]
    {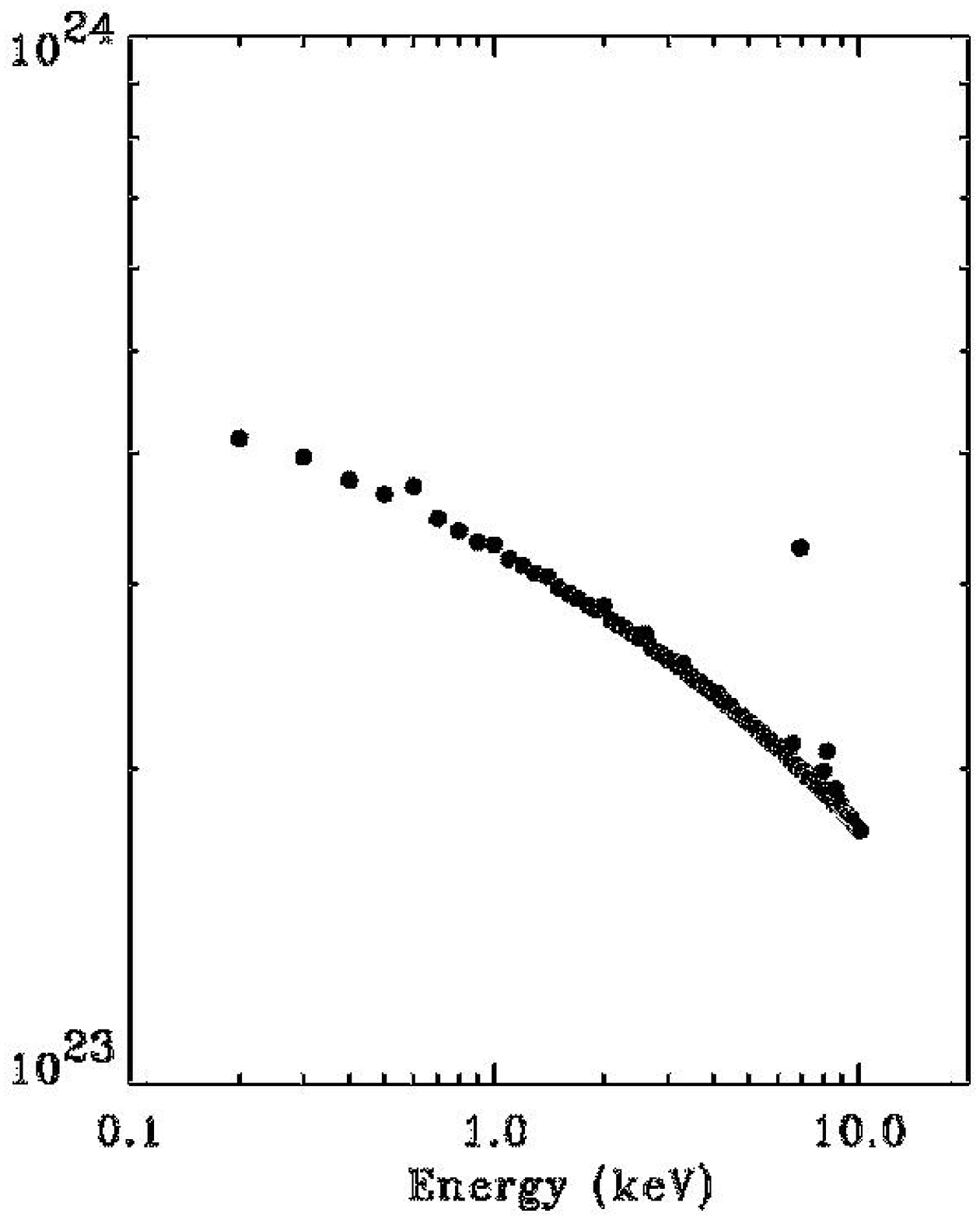}
   \caption{{X-ray luminosities of jet-tori of 63 volume elements 
    with different temperatures: $T$ = $10^{6.64}$\,K, $T$ = $10^7$\,K,
    $T$ = $10^{8}$\,K, $T$ = $10^9$\,K (from {\it left} to {\it right}).
    The jet mass flow rate considered here is
    $ \dot{M}_{\rm j} = 10^{-10}\,{\rm M}_{\odot}{\rm yr}^{-1}$ for
    a $5\,{\rm M}_{\odot}$ central object. 
   }}
   \end{figure*}


\section{X-ray spectra in the rest frame of the volumes}
The computation of the continuum spectrum and the emission lines 
of an optically thin plasma takes into account 
free-free, free-bound and two-photon processes 
(Mewe et al.~\cite{mewe}; Kotani et al.~\cite{kotani}; BK00).
Cosmic abundances given by Allen (\cite{allen}) are used
for a plasma in equilibrium at the local temperature.

\subsection{Luminosities of the fast flow ($T = 10^{6.2}-10^9$\,K)}
Considering the size, density and temperature of each volume,
the luminosities (${\rm erg\,s^{-1} (0.1\,keV)^{-1}}$) 
of the jet-tori have been calculated
in the energy range $0.2-10.1$\,keV (bin size 0.1\,keV).
Examples are shown for four temperatures in Fig.~2
(see also Tab.~1).
With the increase of the tem\-pera\-ture the luminosity range is 
compressed,
therefore those spectra are flatter
and the strong cutoff seen for lower temperatures disappears.
The luminosity of hot gas volume elements ($T \simeq 10^9$\,K),
located above the injection point, is higher (factor 100) 
than the one of the cooler, but faster volume elements.
Note that the luminosities shown in Fig.~2 are calculated
for $\dot{M}_{\rm j} = 10^{-10}\,{\rm M}_{\odot}{\rm yr}^{-1}$.
This quantity is hardly known from observations and, in turn,
the calculated luminosities may constrain its value.
A mass flow rate 100 times higher increases the luminosity by 
a factor of $10^4$, for the same magnetic field geometry. 

For temperatures $T = 10^6-10^9$\,K many emission lines are present 
in the energy range $0.2-10.1$\,keV
(Mewe et al.~\cite{mewe}).
The $0.5-0.9$\,keV band contains O, N, Fe, Ne, S, Ca lines
whereas lines of Ne, Fe, Mg, Ni, Si, S, Ar, Ca are found
between $1.0-4.0$\,keV.
From $6.6-7.0$\,keV mostly FeXXV (He-like) 
and FeXXVI (H-like) emission lines are present
(BK00). 
For fully ionized plasma of $T \ge 10^9$\,K
the bremsstrahlung continuum emission is dominant.

The total rest-frame spectrum (neglecting the velocity of the volumes) 
of a conical sheet of the jet is the integrated luminosity of the
single volumes along the field, taking into account also the number of
volumes along the jet-tori (Fig.~3c).
The emission lines at 6.6 and 6.9\,keV can be identified
as $K\alpha$ lines from He-like and H-like iron, while the one
at 8.2\,keV could be the $K\beta$ from the He-like iron. 

\subsection{The hot flow close to the disk ($T \geq 10^9$\,K)}
The thermal continuum of an optically thin fully ionized plasma follows
from the formula of bremsstrahlung emission 
(Rybicki \& Lightman \cite{rl79}),
\begin{equation} 
\varepsilon_\nu \equiv \frac{dW}{dVdt~d\nu} = 6.8 \times 10^{-38} Z^2
n_e n_i T^{-1/2}e^{-h\nu/kT} \bar{g},
\end{equation}
(in ${\rm erg\,s^{-1}\,cm^{-3}\,Hz^{-1}}$),
with the velocity averaged Gaunt factor $\bar{g}(T, \nu)$\footnote{For 
simplicity, the estimates in this subsection are obtained for a Gaunt 
factor set to unity. 
For the spectra shown in our paper this factor differs slightly from volume
to volume.}, 
the atomic number $Z$, the electron and ion number densities $n_e$ 
and $n_i$,
the Planck constant $h$ and the Boltzmann constant $k$.
Considering the calculated volume parameters
for temperatures below $10^9$\,K,
we obtain a bremsstrahlung luminosity $L_{\rm br}$
comparable to the results in Fig.~2
as expected, in fact, since bremsstrahlung is included in that
calculation.
Still considering $\dot{M}_{\rm j}=10^{-10}\,{\rm M}_{\odot}{\rm yr}^{-1}$,
for $T \simeq 10^{10}$\,K we obtain 
$L_{\rm br} \approx 10^{25}{\rm erg\,s^{-1}(0.1\,keV)^{-1}}$, 
for $T \simeq 10^{11}$\,K 
we obtain $L_{\rm br} \approx 10^{27}{\rm erg\,s^{-1}(0.1\,keV)^{-1}}$,
and when $T \simeq 10^{12}$\,K the luminosity is 
$L_{\rm br} \approx 10^{30}{\rm erg\,s^{-1}(0.1\,keV)^{-1}}$.
Therefore, we expect an increase of the X-ray luminosity due to
the bremsstrahlung contribution of the hottest regions in the jet-disk
system, 
if the optically thin condition is still satisfied there.

   \begin{figure*}
   \centering
    \includegraphics[width=7.2cm]
    {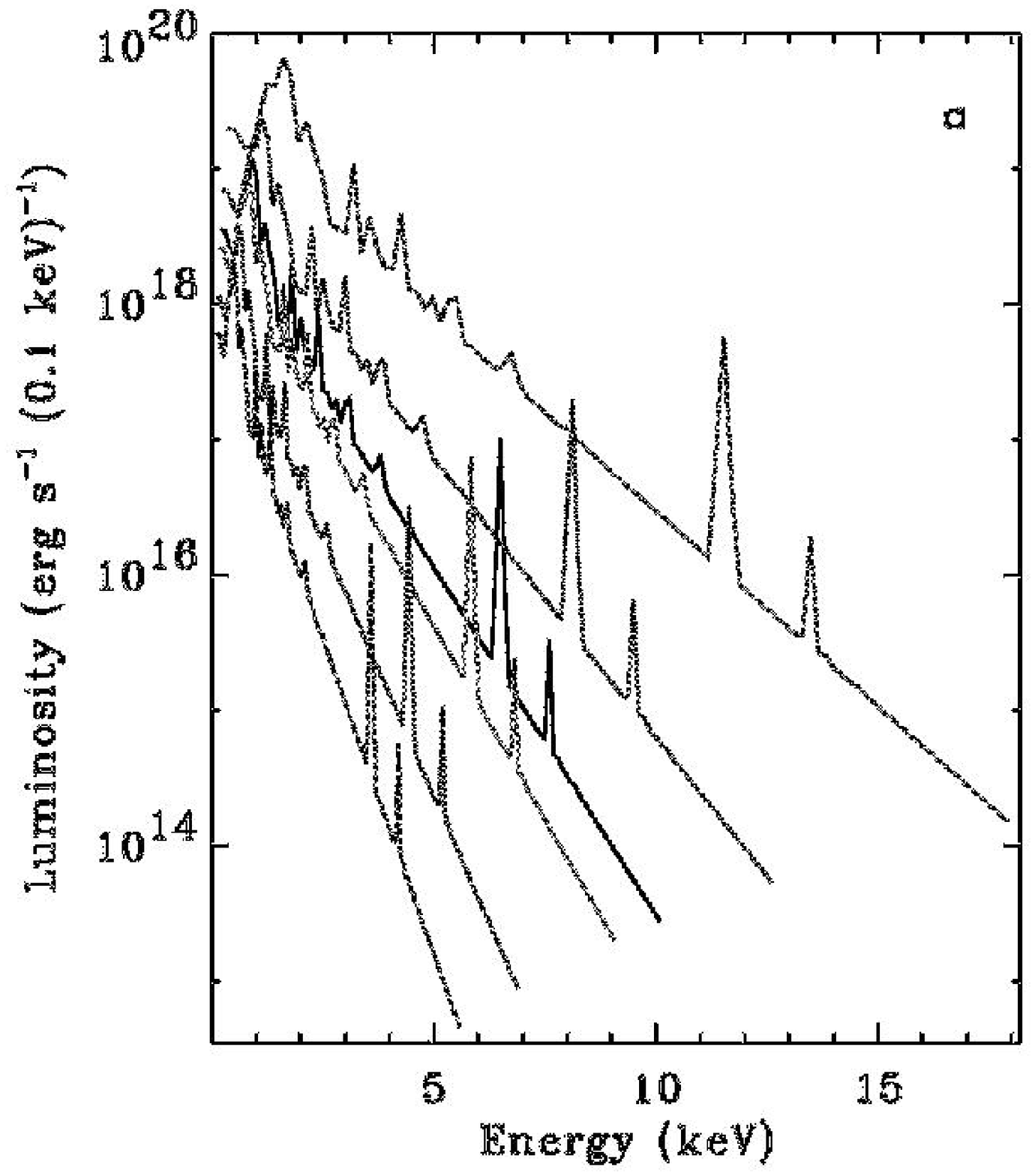}
   \includegraphics[width=7.2cm]
    {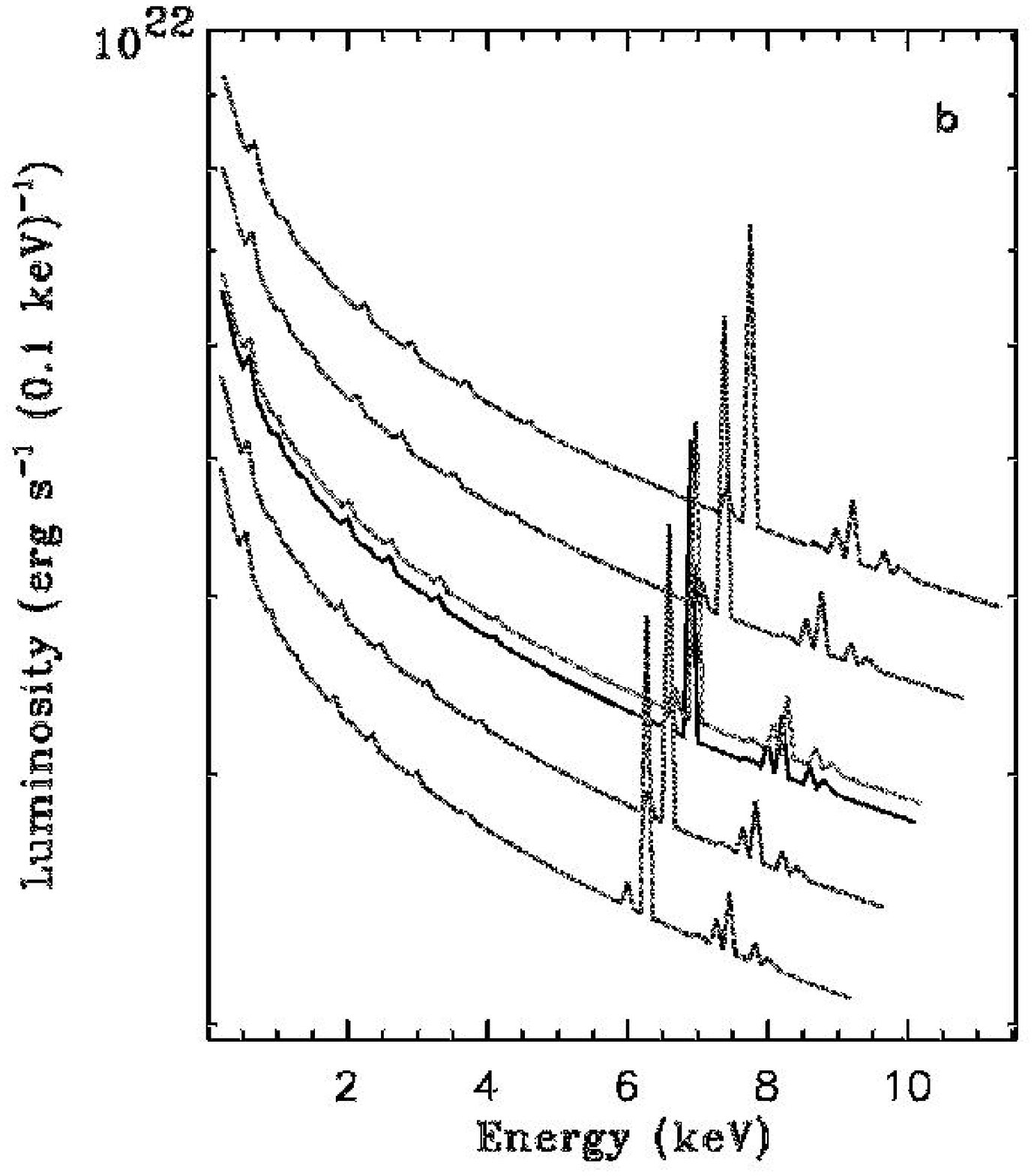}
   \includegraphics[width=7.2cm]
    {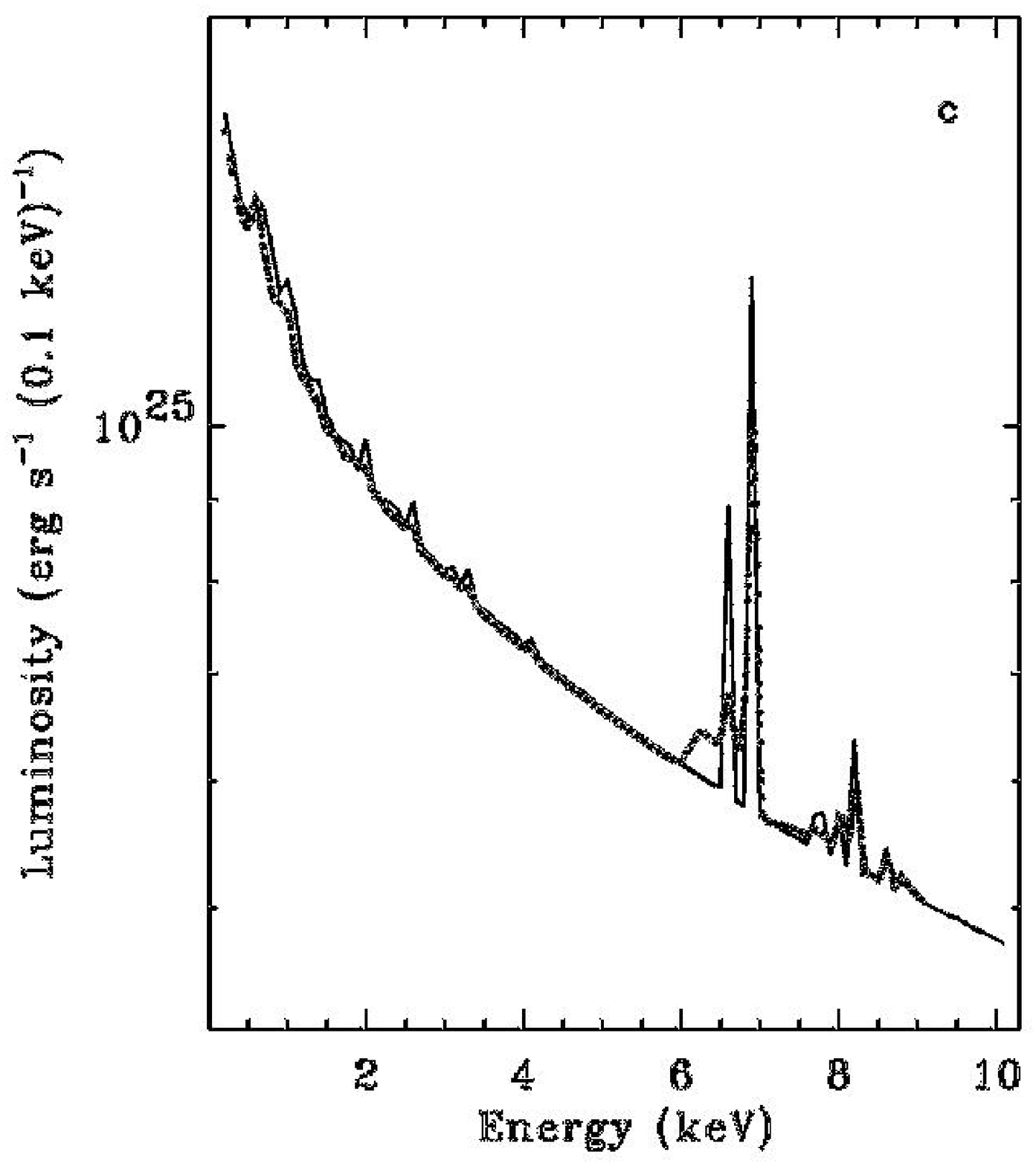}
   \includegraphics[width=7.2cm]
   {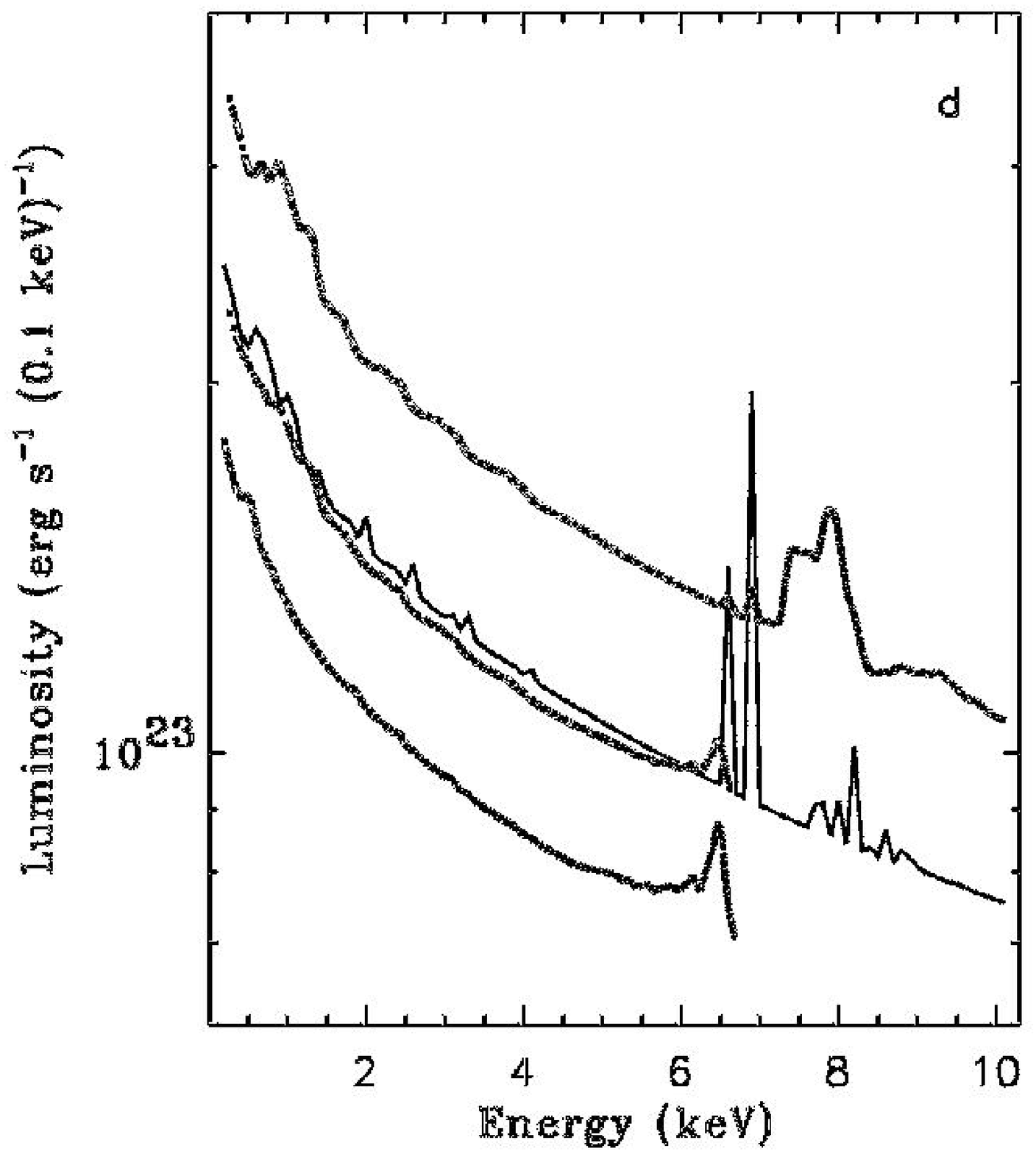}
   \caption{
   Doppler shifted and boosted spectra.
   Spectra for a volume element 
   with $T = 10^7$\,K {\bf a)} and $T = 10^9$\,K
   {\bf b)}
   for different jet inclinations.
   Doppler factor 
   $D_{-40}$ (triple dotted-dashed line, {\it top}),
   $D_{-20}$ (dotted-dashed line, {\it top}), 
   rest-frame (solid line), 
   $D_{\parallel}$ (dotted line),   
   $D_{+20}$ (dotted-dashed line, {\it bottom}), 
   $D_{+40}$ (triple dotted-dashed line, {\it bottom}).
   {\bf c)}
   Comparison of the total shifted/boosted spectrum $D_{\parallel}$ 
   (thick dotted line) of a conical sheet 
   with the rest-frame spectrum (thin solid line).
   {\bf d)}
   Inclined jet,
   comparison of the boosted spectra 
   $D_{-20}$ (thick line {\it top}) and 
   $D_{+20}$ (thick line {\it bottom})  with the 
   total spectrum (thick line {\it middle}) and 
   the rest-frame spectrum (thin line {\it middle}).
   }
   \end{figure*}

\section{Relativistic effects -- Doppler shift and boosting}
We now consider relativistic Doppler effects due to the motion of 
the jet volumes toward the observer. 
The re\-la\-ti\-vis\-tic Doppler factor is 
$
D = (\gamma(1 - \beta \cos\theta))^{-1},
$
where $\gamma$ is the Lorentz factor, $\beta$
the plasma velocity in units of the speed of light and $\theta$ the angle
between the trajectory of the volume and the l.o.s.
The observed energies $E_o$ and  
luminosities $L_o$ of {\em each} volume element are 
shifted and boosted to the rest frame values (index $e$), 
\begin{equation}
E_o = D\,E_e \quad {\rm and} \quad  L_o(E_o) = D^3 \, L_e(E_e).
\end{equation}
Note that the Doppler factor depends on both $\gamma$ and $\theta$ 
(see also Urry \& Padovani \cite{urry})
and is unity for 
$\theta = \arccos \left(\sqrt{(\gamma-1)/(\gamma+1)}\right)$.
For larger angles, relativistic {\em de-amplification} takes place
due to the time lapse in the moving frame of reference.
This is the reason for $D<1$ in our jet (Tab.~1).
Also known as second order Doppler effect, this was first 
observed in SS 433 (Margon \cite{margon}).
De-boosting is also present in the asymptotic radio jets
(different from the collimation region investigated here) 
of GRS\,1915+105 inclined by $70\degr$ to the l.o.s.,
actually providing a distance indicator 
(Mirabel \& Rodr$\acute{\rm {\i}}$guez \cite{MR}).

\subsection{Shifted and boosted spectra}
Figures 3a,b show the effect of boosting and shifting
of the rest frame spectra.
For an angle between the l.o.s. and the jet axis of $40\degr$,
the maximum boosting $D^3 = D_{-40}^3 = 6.7$ is obtained 
for the volume with $T = 10^{6.64}$\,K (see also Tab.~1).
The maximum de-boosting is for the volume at the opposite side of the cone,
$D_{+40}^3 = 0.15$.
As in the rest frame, the 'hot' spectra are flatter.

To obtain a total shifted and boosted spectrum 
we need to interpolate the single volume luminosity values 
since they are shifted to different energies.
Considering the case where the jet axis is along the l.o.s.
($\theta_{\parallel}$, see Fig.~3c),
we have only a weak effect of shifting, 
in fact, we are looking almost perpendicular to an
uncollimated flow. 
For a larger jet inclination the Doppler effects become larger.
In this case, one should take into account the fact that the angle
between the velocity and the l.o.s. ($\theta_{\parallel}$) varies
along the jet-torus.
However, 
we have considered it reasonable to divide the jet-tori in two regions,
one third containing volume elements for which the Doppler
effect has been calculated using the minimum angle 
between the plasma velocity and the l.o.s., and two thirds 
containing volume elements for which the Doppler
effect has been calculated using the maximum angle
between the plasma velocity and the l.o.s.

The total spectra 
have been calculated by first consi\-de\-ring the blue-shifted and
red-shifted parts of the flow and then summing up all the luminosities 
in each energy bin, where blue {\em and} red shifted luminosities
are available.
The result is shown in Fig.~3d with the luminosity rescaled in order 
to compare the total spectrum with its components.

Note that the iron line features are considerably shifted also after
the interpolation.
The change in the line shape is due to the fact that for each
of the 5000 volumes along the jet a different Doppler factor must be 
considered. 
For a larger jet inclination ($D_{-40}$ , $D_{+40}$) the lines are
spread out widely because of the larger Doppler shift (not shown).
The de-boosting contribution of the receding counter-jet 
has not been taken into account. 

\begin{table}
\caption[]{
 Dynamical parameters for four example volume elements.
 Quoted are temperature $T$, mass $M$, particle density $\rho$ 
 and the Lorentz factor $\gamma$.
 The angle $\theta_{\parallel}$ is the angle between the plasma
 velocity and the l.o.s., {\em if} the l.o.s. is parallel 
 to the jet axis.
 The corresponding Doppler factor is $D_{\parallel}$. 
 If the l.o.s. is inclined $20^{\circ}$ to the jet axis, 
 the minimum (maximum) angle between the plasma velocity and the 
 l.o.s. is $\theta_{\parallel} - 20^{\circ}$ 
($\theta_{\parallel} + 20^{\circ}$)
 with a corresponding Doppler factor $ D_{-20} $ ($D_{+20}$)
 and 
similarly for an inclination of $40^{\circ}$. 
}
\begin{flushleft}
\begin{tabular}{|c|c|c|c|c|c|c|c|c|c|}

\hline  
& & & & \vspace{-0.3cm} \\
$T$ (K) & $ 10^9 $  &  $ 10^8 $  & $ 10^7 $  & $ 10^{6.64} $ \\
& & & &  \vspace{-0.3cm} \\
\hline 
& & & & \vspace{-0.3cm}  \\
$M$ (gr) & $ 7  \times 10^7$ & $ 2 \times 10^7$  & $1.1 \times 10^7$  
& $ 0.97 \times 10^7$ \\
$\rho$ (cm$^{-3}$) & $ 6 \times 10^{13}$ & $ 2 \times 10^{12}$ 
& $ 6 \times 10^{11}$ & $ 2 \times 10^{11}$ \\
$\gamma$ & 1.014 & 1.179 & 1.428 & 1.494 \\
$\theta_{\parallel}$ $(^\circ)$  & 82 & 77 & 72 & 70 \\
$ D_{\parallel}$ & 1.010 & 0.960 & 0.898 & 0.899 \\
$ D_{-20} $ & 1.07 & 1.19 & 1.25 & 1.28 \\
$ D_{+20} $ & 0.96 & 0.79 & 0.68 & 0.67 \\
$ D_{-40} $ & 1.12 & 1.47 & 1.77 & 1.88 \\
$ D_{+40} $ & 0.91 & 0.68 & 0.55 & 0.53 \\
\hline

\end{tabular}
\end{flushleft}
\end{table}

\section{Discussion}

\subsection{X-ray luminosities}
We find a total rest-frame X-ray luminosity of the jet 
$ L_{\rm X} = 3.8 \times 10^{31}$ 
$(\dot{M}_{\rm j}/10^{-8}\,{\rm M}_{\odot}{\rm yr}^{-1}){\rm erg\,s^{-1}}$.
The total kinematic luminosity for this jet mass flow rate is
$L_{\rm k} = \gamma \dot{M}_{\rm j} c^2 \approx 10^{39}{\rm erg\,s^{-1}}$
$ >> L_{\rm X} $. 
This proves {\em a posteriori} that the assumption
of a polytropic gas law used to obtain the MHD wind solution
is consistent with the amount of radiation losses.

Considering the Doppler factor $D_{\parallel}$,
the total X-ray luminosity of the jet 
is $L_{\rm X} = 6.4 \times 10^{32}$ 
($\dot{M}_{\rm j}/10^{-8}\,{\rm M}_{\odot}{\rm yr}^{-1}){\rm erg\,s^{-1}}$.
In the case of an inclined jet axis 
($D_{-20}$, $D_{+20}$)
we have $L_{\rm X} = 1.4 \times 10^{33}$ 
($\dot{M}_{\rm j}/10^{-8}\,{\rm M}_{\odot}{\rm yr}^{-1}){\rm erg\,s^{-1}}$.
For $D_{-40}$ and $D_{+40}$ we obtain 
$L_{\rm X} = 1.1 \times 10^{33}$ 
($\dot{M}_{\rm j}/10^{-8}\,{\rm M}_{\odot}{\rm yr}^{-1}){\rm erg\,s^{-1}}$.
These values can be increased by the contribution of
bremsstrahlung from the high temperature 
($T \geq 10^9$\,K)
volumes 
till about $L_{\rm X} \approx 10^{34}{\rm erg\,s^{-1}}$.

In comparison, the X-ray luminosity of GRS\,1915+105 is
$10^{38}{\rm erg\,s^{-1}}$ in low-state and 
$10^{39}{\rm erg\,s^{-1}}$ in high-state
(Greiner et al.~\cite{grein}), 
and larger than the one we obtain.
Such a luminosity might be obtained from the jet for an increased 
mass flow rate. 
The jet inclination of $70\degr$ 
implies a maximum boosting of about 20 for some volumes. 
Further, also the accretion disk contributes to the X-ray flux.
In SS433 we have $ L_{\rm X} > 10^{35}{\rm erg\,s^{-1}}$ 
(Brinkmann et al.~\cite{brink2})
but no broad Fe-lines are observed.
This might be either due to a very low mass flow rate (low jet luminosity)
or to a very high mass flow rate (self-absorption of the emission lines).
%

Higher jet velocities ($\gamma >2$) may increase the Doppler boosting.
Such velocities can be easily obtained for a higher flow magnetization,
i.e.~for a stronger magnetic field strength or a lower jet mass flow rate
(see FG01; Fendt \& Camenzind \cite{fc96}).
However, for the same mass flow rate, a higher velocity implies a lower
gas density, which may lead, instead, to a decrease of the lumino\-sity. 
The interplay of these effects is rather complex.
The {\em rest frame} emissivity depends on the density as $\sim \rho^2$ 
and is also proportional to the emitting volume.
The maximum {\em Doppler boosting} 
increases with the Lorentz factor,
$D^3 (\cos\theta = 1) 
\simeq (2\gamma^2\,(1 + \sqrt{1 - \gamma^{-2}}))^{3/2} $, 
whereas the 
real boosting parameter also depends on the inclination of
the velocity vector to the l.o.s.
Answering
the question how these effects determine the observed X-ray luminosity,
would require a detailed 
study of various MHD wind 
solutions and their derived spectra
investigating different magnetic field geometries (degree of collimation), 
jet mass flow 
rates (the flow magnetization),
and also 
possible masses of the central black hole.
We will return to this important point in a future paper.

Markoff et al.~(\cite{markoff}) have recently shown (for XTE J1118+480) 
that synchrotron emission from the jet may play a role also in the 
X-ray band. 
Their model differs from ours in some respects, especially the initial
jet acceleration is not treated and the jet nozzle geometry is more
concentrated along the axis with a jet radius of only 10 
Schwarzschild radii
(in our model the jet is much wider and collimates later).
As a consequence, the densities become higher and it is questionable 
whether a more reasonable jet geometry will deliver the same amount of
X-ray flux.

\subsection{Jet plasma composition}
At this point we should note that the fundamental question of the
plasma composition in relativistic jets has not yet been answered.
In the case of microquasars 
we do not really
know whether the jet consists of a $e^-p^+$ or a $e^-e^+$ plasma 
(see e.g.~Fender et al.~\cite{fend00}).
It could be possible that these jets are ``light'' jets,
i.e.~made of a pair plasma only, and we would not expect to
observe an iron line emission from such jets. 
Instead, the iron line emission would then arise from processes
connected to the accretion disk or an accretion column.
Such models were discussed for example in the case of XTE\,J1748$-$288 
(Kotani et al.~\cite{kota00}; Miller et al.~\cite{miller}).

On the other hand, the theoretical spectra derived in our paper 
provide an additional information needed in order to interpret the
observed emission lines.
A deeper understanding will, however, require a more detailed
investigation of different jet geometries, viewing angles and mass
flow rates. 
In the end, this might answer the question whether the line emission,
or at least part of it, 
comes from the highly relativistic jet motion or from a rapidly rotating
(i.e.~also relativistic) accretion disk.
For example, we expect the emission lines of a collimated jet being
narrower, and probably shifted by a larger Doppler factor, due to the 
strong beaming.
One should also keep in mind that the direction of motion of the jet
material is inclined (if not perpendicular) to the disk rotation.

Evidently, if the observations would tell us that the Doppler shifted 
Fe lines which are visible in our theoretical spectra arise in the jet
material, this would also prove the existence of a baryonic component
in these jets.

Nevertheless, observations in the radio and shorter wavelengths give 
clear indication for synchrotron emission from highly relativistic 
electrons. 
Whether this non thermal particle population contributes to all of
the observed emission is not clear, a hot thermal plasma may also exist
besides the non thermal electrons. 

A similar discussion concerning the plasma composition is present
in the context of extragalactic jets (e.g.~Mukherjee et al.~\cite{mukh}).
The non thermal emission from blazars can be explained by inverse
Compton scattering of low-energy photons by the relativistic electrons 
in the jet. 
However, two main issues remain unsolved: the source of the soft photons
that are inverse Compton scattered, and the structure of the inner jet,
which cannot be imaged directly. 
The soft photons can originate as synchrotron emission either within
the jet (see e.g.~Bloom \& Marscher \cite{bloom}) 
or nearby the accretion disk, 
or they can be disk radiation reprocessed in broad emission line clouds 
(see e.g.~Ghisellini \& Madau \cite{ghisellini}). 
In contrast to these leptonic jet models, the proton-initiated cascade
model (see e.g.~Mannheim \& Biermann \cite{mannheim}) 
predicts that the high-energy
emission comes from knots in jets as a consequence of diffusive shock
acceleration of protons to energies so high that the threshold of
secondary particle production is exceeded.

Comparison of our calculated Fe emission lines to the observed ones 
potentially give some hints on the plasma composition 
($e^-p^+$ or $e^-e^+$) in relativistic jets.

\section{Summary}
For the first time, theoretical thermal X-ray spectra were obtained 
for the dynamical parameters of a relativistic jet calculated
from the MHD wind equation.
The total spectra were derived as composition of the spectral contributions
of the single volume elements accelerating along the jet with relativistic 
speed. 
Our results are the following.

\begin{enumerate}

\item We find X-ray emission from the hot inner part of the jet
originating in a region of $2.5 \times 10^{-5}$\,AU diameter close
to the center of a 5\,${\rm M}_{\odot}$ jet source.
The jet X-ray luminosity is $ L_{\rm X} \sim 10^{33}$ 
($\dot{M}_{\rm j}/10^{-8}\,{\rm M}_{\odot}{\rm yr}^{-1}){\rm erg\,s^{-1}}$.

\item
Emission lines of Fe~XXV and Fe~XXVI are clearly visible in our spectra. 
Interestingly, the $K\!\alpha$ iron emission line
has been probably observed in GRS 1915+105 
(Ebisawa et al.~\cite{ebisawa}) and 
XTE J1748$-$288 (Kotani et al.~\cite{kota00}).
The absence of broad Fe-lines in the spectrum of SS433 might tell us 
something
on the 'invisibility' of the acceleration region above the disk.
Comparison of our calculated emission lines to observed ones 
may give some hints on the plasma composition 
in relativistic jets.

\item 
From the MHD jet underlying the spectra we find a maximum Doppler boosting
of about 7. 
Minimum boosting is present along the opposite side of the jet cone
(Doppler factor 0.53).
The shift of the emission lines is always visible. 
The boosting, however, does not play a major role in the total spectra, 
because of the uncollimated geometry of the innermost part of the
jet emitting the X-rays and the combined effect of 
boosting and de-boosting around the jet cone.
 
\end{enumerate}

\noindent
If jets from X-ray binaries indeed contain matter of baryonic composition,
our model will have a broad application.
Indication of that is probably given by the observation of iron emisson 
lines in some sources (see above).
However, it is not yet clear, whether the line emission originates in 
the jet or in the accretion disk.
Our calculated Fe emission lines may help to interpret the observed spectra
and potentially give some clue on the plasma composition in 
relativistic jets.

This study will be extended in a future work
investigating spectra of jets with different magnetic geometry,
mass flow rates and central masses.
In the end, this might also allow to constrain the intrinsic parameters of
jet formation itself (such as mass loading or opening angle) from the
observation of the large-scale, asymptotic jet.

\begin{acknowledgements}
  This work was partly supported by the German Science Foundation 
  (Deut\-sche For\-schungs\-ge\-mein\-schaft) as project DFG/FE490.
  We thank an anonymous referee for useful comments.
\end{acknowledgements}

\end{document}